\documentclass[a4paper,12pt]{article}
\pdfoutput=1
\usepackage{graphicx, rotating,amssymb,amsmath}

\ifx\pdfoutput\undefined
\usepackage[bookmarks]{hyperref}	
\else
\usepackage{hyperref}	
\fi
\hypersetup{colorlinks,bookmarksopen,bookmarksnumbered,citecolor=verde,
linkcolor=blus,pdfstartview=FitH,urlcolor=rosso}
\def\myurl#1#2{\href{http://#1}{#2}}
\def\hhref#1{\href{http://arxiv.org/abs/#1}{#1}} 

\usepackage{multicol}
\usepackage{color}
\definecolor{rosso}{cmyk}{0,1,1,0.4}
\definecolor{rossos}{cmyk}{0,1,1,0.55}
\definecolor{rossoc}{cmyk}{0,1,1,0.2}
\definecolor{blu}{cmyk}{1,1,0,0.3}
\definecolor{blus}{cmyk}{1,1,0,0.6}
\definecolor{bluc}{cmyk}{1,1,0,0.1}
\definecolor{verde}{cmyk}{0.92,0,0.59,0.25}
\definecolor{verdec}{cmyk}{0.92,0,0.59,0.15}
\definecolor{verdes}{cmyk}{0.92,0,0.59,0.4}

\def\Fermi{{\sc Fermi}} 
\def\PAMELA{{\sc Pamela}}

\begin{document}

\begin{flushleft}
\tiny{CERN-PH-TH/2012-214 \hfill 
SACLAY--T12/054 \hfill
LAPTH-040/12 \hfill
}
\end{flushleft}

\vspace{0.2cm}

\begin{center}


{\LARGE \bf \PAMELA\ and \Fermi\ limits\\[0.4cm] on the neutralino-chargino mass degeneracy}

\vspace{1cm}

{
{\bf\large Genevi\`eve B\'elanger}$^{\,a}$,
{\bf\large C\'eline B\oe hm}$^{\,b,a}$, \\[2mm]
{\bf\large Marco Cirelli}$^{\,c,d}$,
{\bf\large Jonathan Da Silva}$^{\,a}$, 
{\bf\large Alexander Pukhov}$^{\,e}$
}
\\[7mm]
{\it $^a$ \href{http://lapth.in2p3.fr/}{LAPTh}, U. de Savoie, CNRS, BP 110, Annecy-le-Vieux F-74941, France}\\[3mm]
{\it $^b$ \href{http://www.ippp.dur.ac.uk}{Institute for Particle Physics Phenomenology},\\ University of Durham, Durham, DH1 3LE, UK}\\[3mm]
{\it $^c$ \href{http://ph-dep-th.web.cern.ch/ph-dep-th/}{CERN Theory Division}, CH-1211 Gen\`eve, Switzerland}\\[3mm]	
{\it $^d$ \href{http://ipht.cea.fr/en/index.php}{Institut de Physique Th\'eorique}, CNRS, URA 2306 \& CEA/Saclay,\\ 
	F-91191 Gif-sur-Yvette, France}\\[3mm]

{\it $^e$ \href{http://www.sinp.msu.ru/eng/sinp.php3}{Skobeltsyn Institute of Nuclear Physics}, Moscow State University,\\ Moscow 119992, Russia}

\end{center}

\date{today}

\centerline{\large\bf Abstract}
\begin{quote}
\color{black}
Searches for Dark Matter (DM) particles with indirect detection techniques have reached important milestones with the precise measurements of the anti-proton ($\bar{p}$) and gamma-ray spectra, notably by the \PAMELA\ and  \Fermi-LAT experiments. While the $\gamma$-ray results have been used to test the thermal Dark Matter hypothesis and constrain the Dark Matter annihilation cross section into Standard Model (SM) particles, the anti-proton flux measured by the \PAMELA \ experiment remains relatively unexploited. Here we show that the latter can be used to set a constraint on the neutralino-chargino mass difference. To illustrate our point  we use a Supersymmetric model in which the gauginos are light, the sfermions are heavy and the Lightest Supersymmetric Particle (LSP) is the neutralino. In this framework the $W^+ W^-$ production is expected to be significant, thus leading to large $\bar{p}$ and $\gamma$-ray fluxes. After determining a generic limit on the Dark Matter pair annihilation cross section into $W^+ W^-$ from the 
$\bar{p}$ data only, we show that one can constrain scenarios in which the neutralino-chargino mass difference is as large as $\simeq$ 20 GeV for a mixed neutralino (and intermediate choices of the $\bar p$ propagation scheme). This result is consistent with the limit obtained by using the \Fermi-LAT data. 
As a result, we can safely rule out the pure wino neutralino hypothesis if it is lighter than 450 GeV and constitutes all the Dark Matter.
\end{quote}

\newpage

\section{Introduction}
\label{introduction}

Indirect searches for Dark Matter, i.e. searches for `anomalous' features in cosmic rays (e.g. gamma-rays, neutrinos, positrons and anti-protons), have been proposed in the late 70's as a powerful way to reveal the existence of Dark Matter  annihilations in the Milky Way halo and beyond~\cite{indirectgamma,indirectcharged}. 
These techniques are meant to give precious insights about the nature of the Dark Matter particle and its properties, assuming that a signal is seen. Yet there are several limiting factors which weaken their ability to elucidate the dark matter problem. In particular indirect detection requires a detailed knowledge of the astrophysical backgrounds and foregrounds and therefore depends on the present knowledge of astrophysical sources. To make a discovery one either has to carefully remove known (or modelled) background in order to expose the `anomalous' component or hope that the Dark Matter signal is well above the background and exhibits very clear features, which would be difficult to mimic by invoking astrophysical sources only.

\medskip

Currently there are contradicting claims regarding whether indirect detection is giving clues of dark matter or not. On one hand, there are  possible anomaly detections which could be explained in terms of Dark Matter annihilation or decay. These include for example the positron excess, as seen in \PAMELA\ (and \Fermi-LAT) data~\cite{PAMELApositrons}, a possible feature in the $e^+ + e^-$ spectrum~\cite{epem}, a claimed $\gamma$-ray excess at $\sim$ 10 GeV energies~\cite{Hooper10GeV}  \footnote{All these claims have possible drawbacks, cf \cite{Boehm:2002yz,Boehm:2010kg,Bertone:2008xr,Cirelli:2009vg,Cirelli:2009bb,Cirelli:2009dv,Cirelli:2012ut}.} and, most recently, two possible $\gamma$-ray lines at 111 and 129 GeV~\cite{line130GeV,noline130GeV}. On the other hand, a large bulk of present astrophysical data essentially seem to validate the modelling of astrophysical background sources in the GeV-TeV range (disregarding these possible anomalies), and therefore enables one to set powerful constraints on the Dark Matter properties. 

\medskip
By measuring the gamma-ray spectrum over a large energy range relevant for Dark Matter physics, the \Fermi-LAT\ collaboration has been able to  set stringent limits on the Dark Matter pair annihilation cross section into Standard Model particles. For example, using the diffuse $\gamma$-ray emission in dwarf spheroidal (dSph) galaxies~\cite{FermiDwarfs} and also in the Milky Way~\cite{FermiDiffuseMW,Ackermann:2012qk}, the \Fermi-LAT\ collaboration has ruled out Dark Matter candidates with a total  annihilation cross section of $\langle \sigma v \rangle = 3 \times  10^{-26} \ \rm{cm^3/s}$ if $m_{\rm DM} \lesssim 30$ GeV. 
This constituted a remarkable milestone as such a value corresponds to that suggested by the thermal freeze-out scenario, which is generally considered as a strong argument in favour of Weakly Interacting Massive Particles (WIMPs).

\medskip

These limits nevertheless weaken at higher DM masses, therefore allowing for heavier DM candidates with a larger pair annihilation cross section. For example, for $m_{\rm{DM}} =$ 100 GeV the limit relaxes to $\langle \sigma v \rangle \lesssim 10^{-25} \ \rm{cm^3/s}$ while for $m_{\rm DM} =$ 500 GeV, it reads $\langle \sigma v \rangle \lesssim 3 \times 10^{-25} \ \rm{cm^3/s}$, which is one order of magnitude higher than the `thermal' cross section.

\medskip
DM models with such large values of the pair annihilation cross section  have actually been proposed over the last five years as a consequence of the excesses in $e^+$ and $e^++e^-$ fluxes. While they may remain hypothetical, discovering such a configuration would invalidate the WIMP `vanilla' model and either point towards the existence of non-thermal process in the Early Universe (possibly opening up an unexpected window on fundamental physics at high energies) or potentially call for more sophisticated mechanisms, such as Freeze-In and regeneration as proposed in \cite{Hall:2009bx,Chu:2011be}. Explaining the observed dark matter relic density may remain nevertheless  challenging. For example, in \cite{Williams:2012pz}, it was shown that candidates with a total annihilation cross section exceeding $\langle \sigma v \rangle =  10^{-24} \ \rm{cm^3/s}$ (corresponding to a thermal relic density smaller than $3 \%$) would be ruled out by the \Fermi-LAT experiment if they were regenerated at 100$\%$. 

\medskip

In addition to measurements of the $e^+$ and $e^+ + e^-$ spectra mentioned above, there is also the measurement of the galactic $\bar{p}$ flux, presented by the \PAMELA\ collaboration~\cite{PAMELApbar1,PAMELApbar2}. While extensive work was done to explain the electron/positron excesses in terms of Dark Matter annihilations (or decays), the implications of the absence of anomalies in the $\bar{p}$ spectrum has remained relatively unexploited. Indeed only a relatively small number of works~\cite{CKRS,salati,boehm,Evoli:2011id,Garny:2011ii,
Asano:2011ik,Garny:2012eb} have dealt with it and shown that large Dark Matter annihilation cross sections can be constrained by the  \PAMELA\ data. Among the most interesting conclusions which have been reached let us cite for example that in \cite{salati} constraints on the annihilation cross section into $b \bar{b}$ were given (for the same mass range as is considered in this paper) and limits on the $W^+ W^- $ final state were mentioned for $m_{\rm DM} = 1$ TeV and one specific set of propagation parameters. In \cite{Asano:2011ik}, constraints on the $q\bar{q}g$ were set for bino-like neutralinos.

\medskip

The first aim of this paper is therefore to propose a more systematic analysis of these general anti-proton constraints on the DM annihilation cross section, including paying attention to the uncertainties associated with DM and astrophysical predictions. The  second aim of the present analysis is to demonstrate that these measurements can actually constrain the properties of specific DM scenarios, including the mass spectrum in the dark sector. To illustrate this, we will work within a `simplified' version of the phenomenological Minimal Supersymmetric Standard Model (pMSSM)~\cite{Djouadi:1998di} in which all sfermion masses are set to 2 TeV, except for the stop and sbottom masses. The soft masses for the stop are allowed to be much lighter to obtain a Higgs at 125 GeV. In this scenario the only particles with masses below the TeV threshold are therefore the neutralino, chargino, the supersymmetric Higgses and the lightest stop and sbottom. Such a configuration of `light' gauginos and heavy sfermions may actually seem unnatural from a supersymmetric point of view (albeit close to split SUSY~\cite{splitSuSy}) but it is supported by the unfruitful searches for squarks and gluinos at LHC, at least to some extent~\footnote{Even though, admittedly, those negative searches may also be a sign that Supersymmetry is not realised at the TeV scale.}. 

\medskip
With this very set up in mind, one can investigate scenarios where the neutralino pair annihilation cross section into $W^+ \, W^-$  gauge bosons is enhanced (due in particular to the chargino exchange diagram). Such a large annihilation cross section gives both a significant anti-proton and diffuse gamma ray flux, together with a gamma ray line, and is therefore  potentially constrained by the \PAMELA\ and \Fermi-LAT data.
In Supersymmetry, such an enhancement is realised when the LSP neutralino is mass degenerated with the chargino, i.e. when the neutralino has a significant wino component. The combination of both \Fermi-LAT and \PAMELA\ data is therefore expected to constrain the wino fraction of the lightest neutralino, thus realizing our second aim.  Note that constraints on the neutralino composition are also expected to be obtained in presence of a lower sfermion mass spectrum. However the effect of the chargino-neutralino mass degeneracy on $\gamma-$ray and $\bar{p}$ production would be much harder to characterise. Hence our choice in favour of a heavy sfermion mass spectrum.

\medskip

The paper is organised as follows. In section \ref{constraints} we derive generic constraints on the Dark Matter pair 
annihilation cross section into $W^+W^-$ from anti-proton data and recall the \Fermi-LAT limits that are obtained from gamma-ray observations in the Milky Way and dwarf Spheroidal  galaxies. In Section \ref{SuSy} we present the Supersymmetric model that we shall consider and explain how we perform the scans of the parameter space. Finally in Section \ref{results} we apply the \PAMELA\ and \Fermi-LAT  limits to our SUSY model and show that the anti-proton data can be more constraining than gamma-ray observations. We conclude in Section \ref{conclusions}.


\section{Anti-proton and $\gamma$-ray bounds on  $\sigma_{{\rm DM}  \ {\rm DM} \  \rightarrow \ W^+ W^-}$}
\label{constraints}

In this section we discuss how anti-proton and gamma ray data impose generic constraints on the Dark Matter pair annihilation cross section into $W^+ W^-$ as a function of the Dark Matter mass.


\subsection{Generic bounds on $\sigma_{{\rm DM} \ {\rm DM} \ \rightarrow \ W^+ W^-}$ from anti-protons}
\label{pbarconstraints}

$W^\pm$ production in space leads to abundant anti-proton production as the $W^\pm$'s decay products hadronize. The flux of anti-protons thus produced by DM annihilations into a pair of $W^{\pm}$ gauge bosons in the Milky Way and collected at Earth is therefore determined by the Dark Matter pair annihilation cross section into $W^+ W^-$, the Dark Matter mass and the Dark Matter halo profile. It also depends on the anti-proton propagation parameters which are being considered. Hereafter we will assume that the dark matter halo profile is well described by an Einasto profile (we checked that other choices make a small difference) and consider the standard three sets of propagation parameters (`MIN', `MED', `MAX') summarised in table~\ref{tab:proparam}. 
In practice, we use the anti-protons fluxes which are given in~\cite{PPPC4DMID}, to which we refer for further details. 

\begin{table}[t]
\center
\begin{tabular}{c|cccc}
 &  \multicolumn{4}{c}{Antiproton parameters}  \\
Model  & $\delta$ & $\mathcal{K}_0$ [kpc$^2$/Myr] & $V_{\rm conv}$ [km/s] & $L$ [kpc]  \\
\hline 
MIN  &  0.85 &  0.0016 & 13.5 & 1 \\
MED &  0.70 &  0.0112 & 12 & 4  \\
MAX  &  0.46 &  0.0765 & 5 & 15 
\end{tabular}
\caption{\em \small {\bfseries Propagation parameters} for anti-protons in the galactic halo (from~\cite{FornengoDec2007,DonatoPRD69}). Here $\delta$ and $\mathcal{K}_0$ are the index and the normalization of the diffusion coefficient, $V_{\rm conv}$ is the velocity of the convective wind and $L$ is the thickness of the diffusive cylinder. 
\label{tab:proparam}}
\end{table}

\smallskip

In order to constrain the annihilation cross section, we will consider that all present data define the maximal flux in anti-proton that is allowed by the PAMELA~\cite{PAMELApbar2} experiment~\footnote{To avoid the uncertainty related to solar modulation, we restrict ourselves to using the \PAMELA\ data above an anti-proton energy of 10 GeV.}. Both the predicted energy spectrum and the flux depend on the dark matter mass that is being assumed. For each value $m_{\rm{DM}}$, we will therefore compare the sum of the astrophysical background flux and predicted anti-protons flux originating from Dark Matter with the \PAMELA\ data. Given the uncertainties on the astrophysical background, we will apply two different procedures to derive meaningful limits.  One can be regarded as aggressive (it assumes a fixed background) while the other one is more conservative (the background can be adjusted within the uncertainties).

\begin{itemize}
\item[$\circ$] For obtaining aggressive limits (referred to as  {\it fixed background} in the following), we adopt the standard flux of astrophysical (secondary) anti-protons from~\cite{BringmannSalati} and add it to the DM anti-protons flux. We then compare the result with the \PAMELA\ data and derive a 95\% C.L. limit by imposing that the global $\chi^2$ of the background $+$ DM flux does not exceed by more than 4 units the $\chi^2$ of the null hypothesis (background only).

\item[$\circ$] For obtaining conservative limits (hereafter referred to as {\it marginalized background}), we take again the standard form of the background spectrum predicted in~\cite{BringmannSalati}, except that now we allow for the normalisation of the background spectrum $A$ and the spectral index $p$ to vary within 40$\%$ and $\pm 0.1$ respectively (for each value of the DM mass and pair annihilation cross section into $W^+ \, W^-$). 

In practice, we multiply the standard description of the background spectrum by a factor $A \, (T/T_0)^p$, where $T$ is the anti-proton kinetic energy, $T_0 = 30$ GeV is a pivot energy and with $0.6 < A < 1.4$ and $-0.1 < p < +0.1$. These are quite generous intervals, which allow to include the uncertainty predicted in~\cite{BringmannSalati}.  We then add up the DM contribution expected for each point in the parameter space defined as $(m_{\rm DM}, \langle \sigma v \rangle)$ and identify the pair of parameters $A$ and $p$ which minimises the global $\chi^2$ with the \PAMELA\ data. This procedure therefore corresponds to marginalising over the parameters of the uncertain astrophysical background point-by-point in the DM parameter space. Again, the 95\% C.L. is then imposed by requiring that the marginalised global $\chi^2$ does not exceed 4 units with respect to the null hypothesis (which has been marginalised consistently). 

By considering a variable background spectrum (within the uncertainties) for each value of the DM mass and cross section, we can increase the gap between the expected $\bar{p}$ background and the actual \PAMELA\ data. As a result this leaves more space for a possible DM injection of anti-protons and leads to weaker limits on the DM pair annihilation cross section. A similar approach was used in~\cite{Delahaye:2011jn} but to reduce the gap between the astrophysical background and the data.
\end{itemize}

\begin{figure}[t]
	\parbox[b]{.49\linewidth}{
		\includegraphics[width=\linewidth]{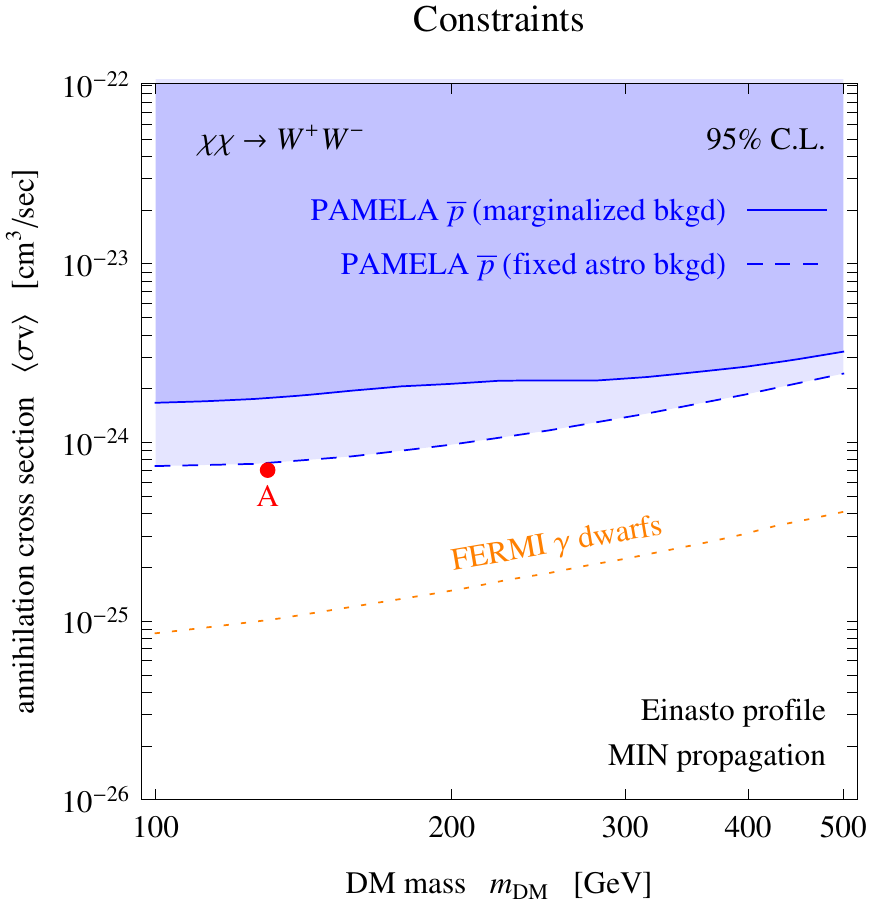}}
	\parbox[b]{.49\linewidth}{ 
		\includegraphics[width=\linewidth]{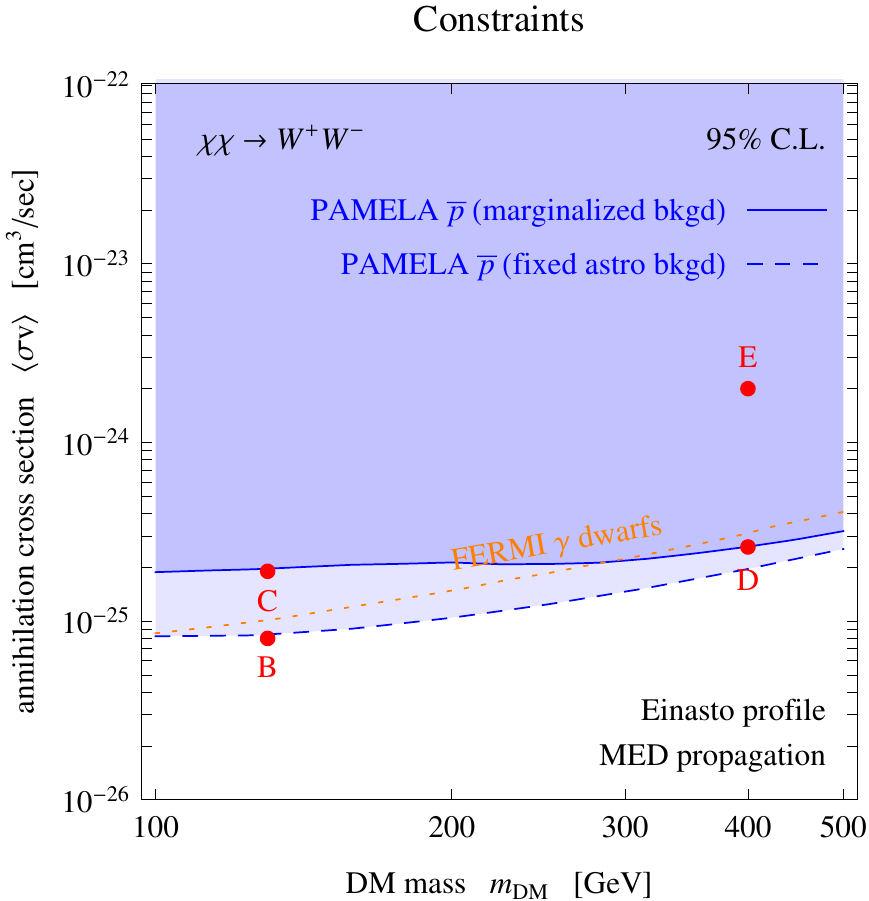}} \\[2mm]
	\parbox[b]{.49\linewidth}{
		\includegraphics[width=\linewidth]{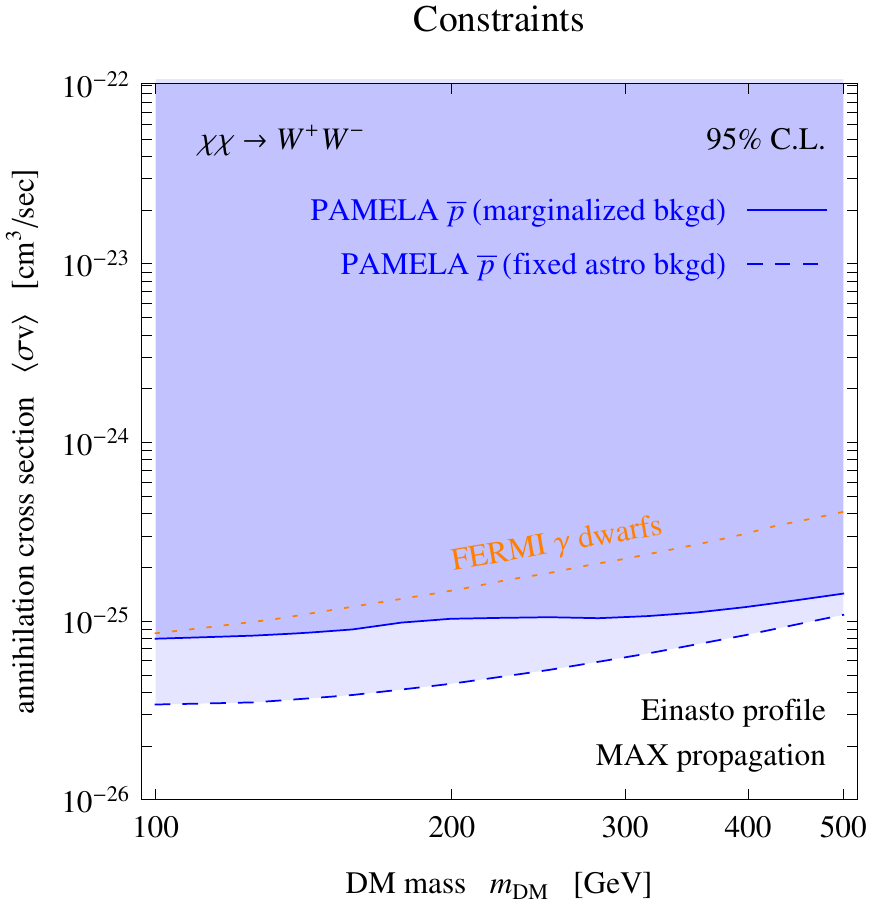}}\hfill
	\parbox[b]{.42\linewidth}{
		\caption{\em \small {\bfseries Anti-proton constraints} on DM annihilation into $W^+W^-$. The upper left, upper right and lower panels refer respectively to the `MIN', `MED' and `MAX' propagation parameters. The constraints obtained by the \Fermi-LAT collaboration \ from satellite dwarf galaxies are superimposed. We also display five benchmark points. {\color{white} Filling. Filling. Filling. Filling. Filling. Filling. Filling. Filling. Filling. Filling. Filling. Filling. Filling. Filling. Filling. Filling. Filling. Filling. Filling. Filling. Filling. Filling. Filling. Filling. Filling. Filling. Filling. Filling. Filling. Filling. Filling. Filling. Filling. Filling.}}
\label{antip_crosssection}
	}
\end{figure}


\medskip

Our constraints are displayed in Fig.~\ref{antip_crosssection} for the `MIN',`MED', `MAX' set of parameters. As expected, the `conservative' limits are slightly less constraining than the `aggressive' ones. 
Also we find that the choice of propagation parameters has a big impact on the type of constraints that can be set: in terms of cross sections, the difference between the `MIN' and `MAX' limits exceeds a factor 10.

\begin{figure}
	\parbox[b]{.32\linewidth}{
		\includegraphics[width=\linewidth]{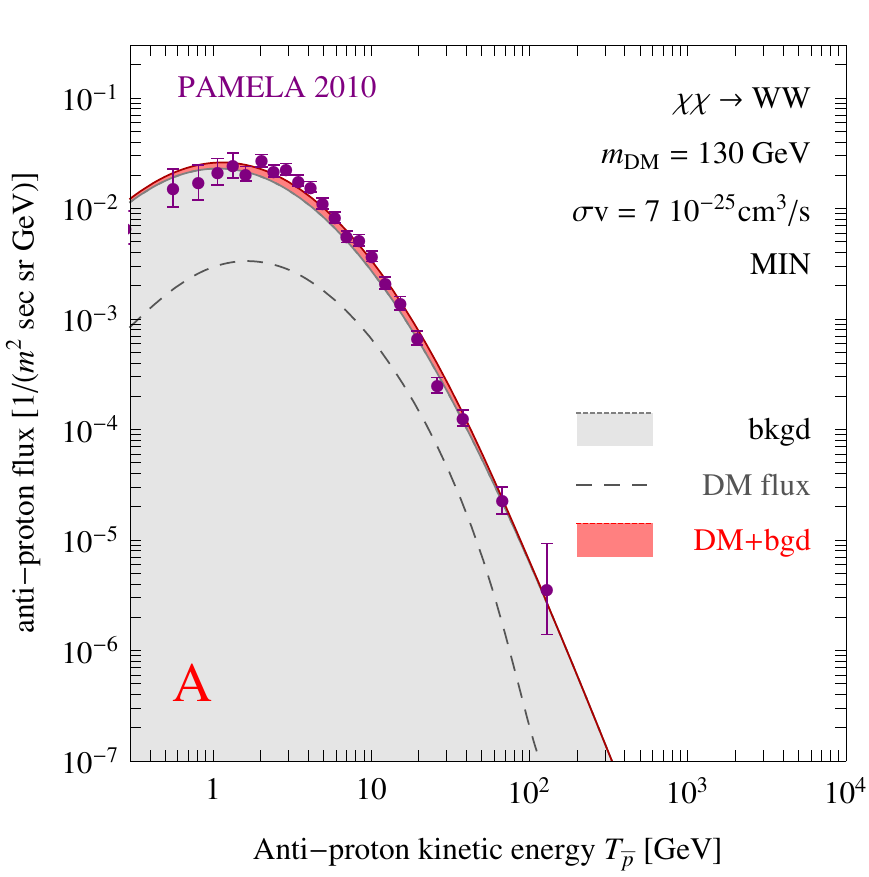}}
	\parbox[b]{.32\linewidth}{ 
		\includegraphics[width=\linewidth]{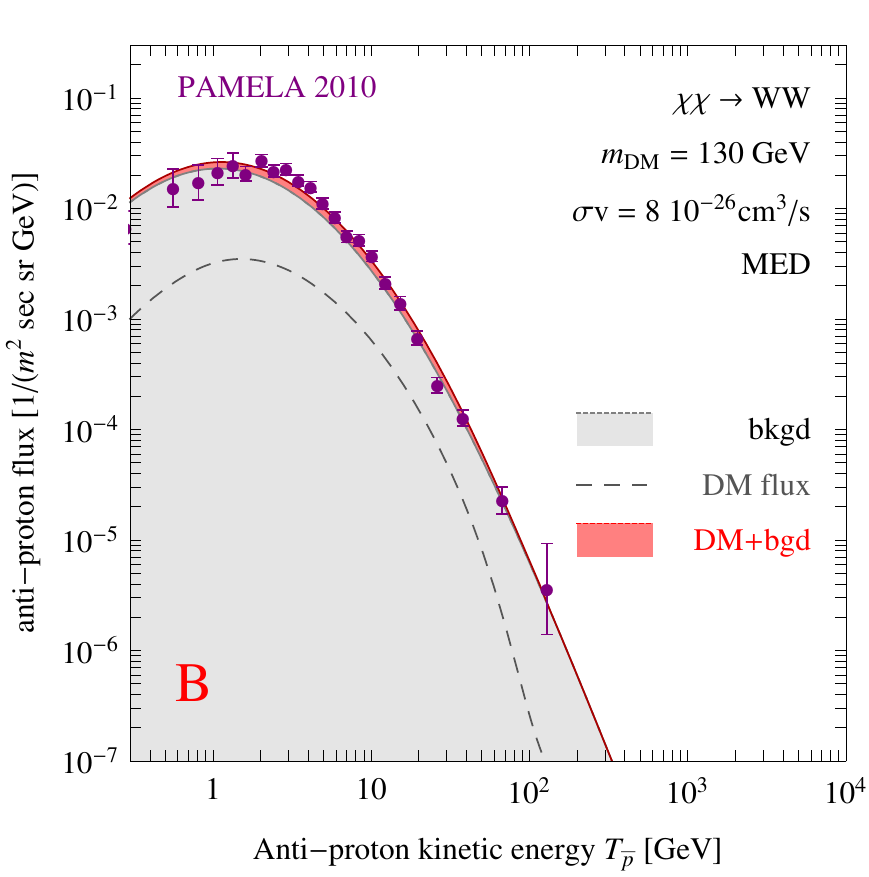}} 
	\parbox[b]{.32\linewidth}{
		\includegraphics[width=\linewidth]{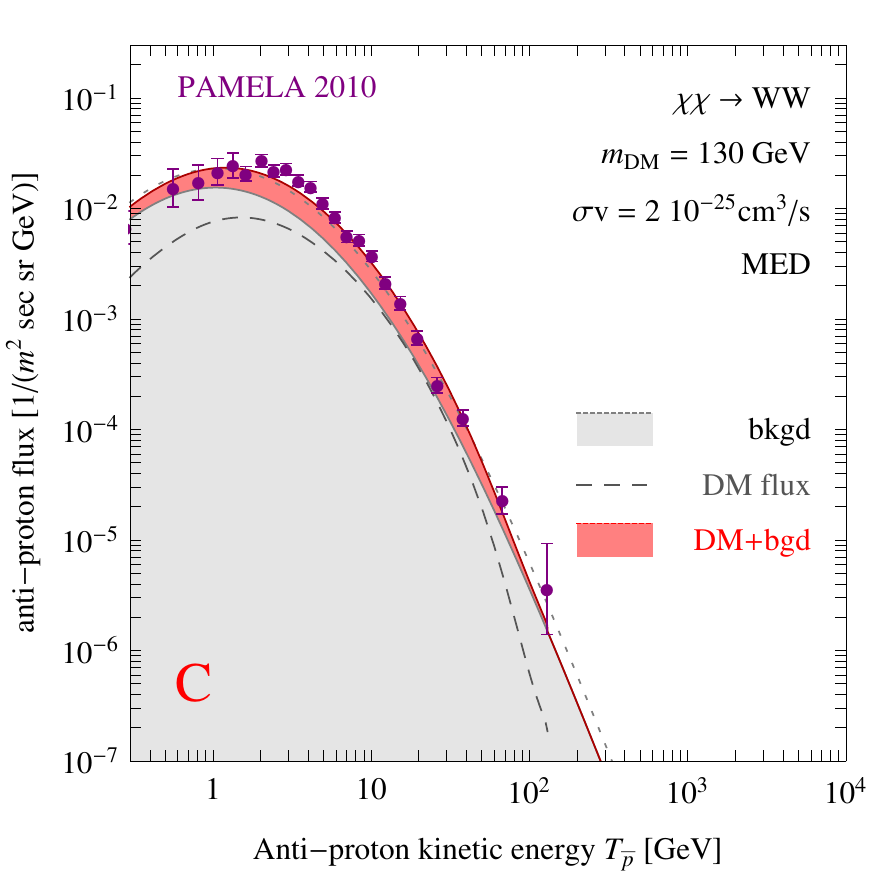}}\\
	\parbox[b]{.32\linewidth}{
		\includegraphics[width=\linewidth]{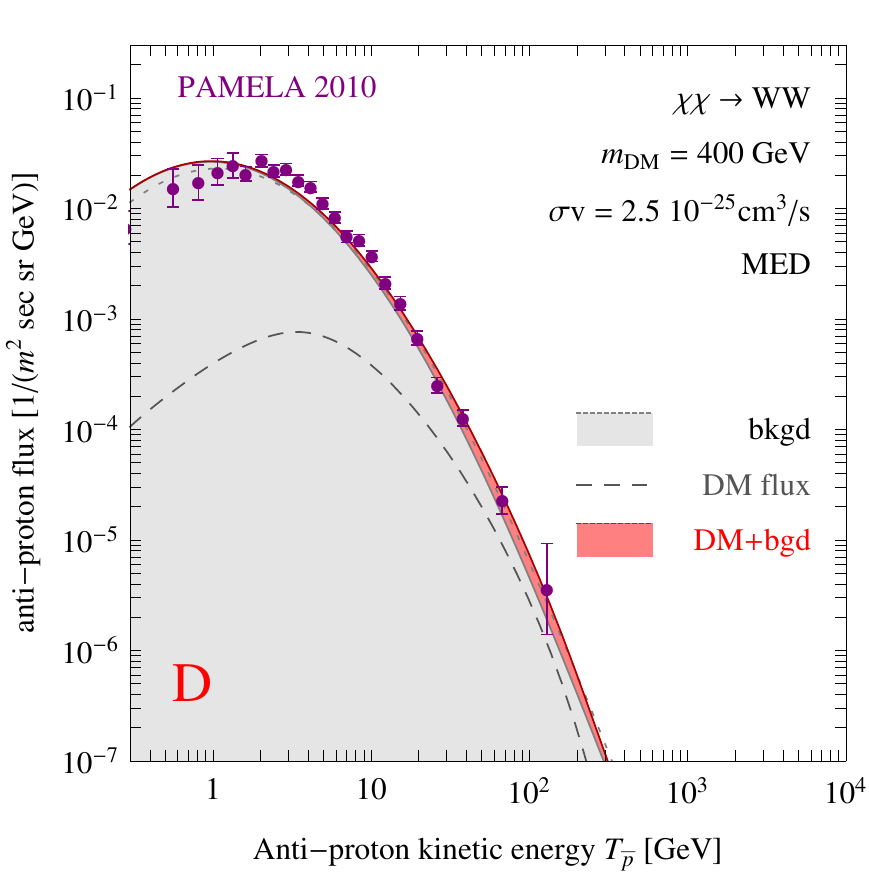}}
	\parbox[b]{.32\linewidth}{
		\includegraphics[width=\linewidth]{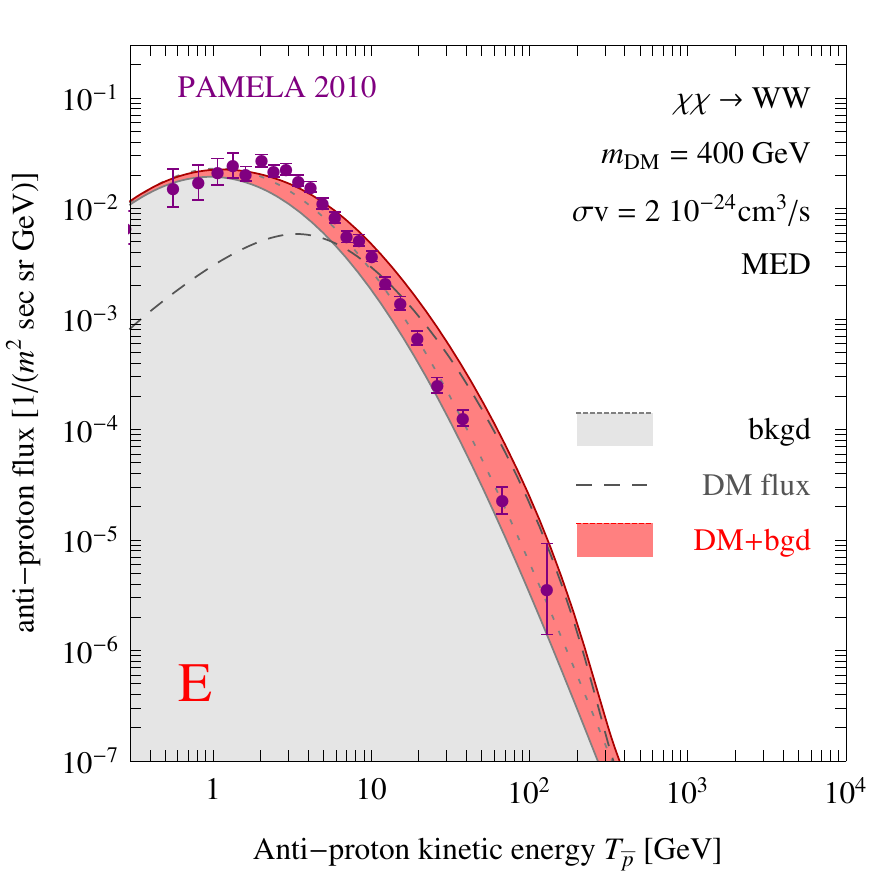}}\hfill
	\parbox[b]{.30\linewidth}{
		\caption{\em \small {\bfseries Examples of the fluxes of anti-protons} (astrophysical background and DM-produced) compared with the data from the \PAMELA\, experiment for the sample points A to E as defined in Fig.~\ref{antip_crosssection}. In each panel the assumed parameters (DM mass, annihilation cross section and propagation scheme) are reported.}
		\label{examples}
	}
\end{figure}

To understand more precisely how these constraints work, we defined 5 scenarios (hereafter referred to as `A',`B',`C',`D',`E'), corresponding to different DM masses, cross sections, propagation parameters and constraint procedures. The corresponding fluxes are plotted in Fig.~\ref{examples}.  As one can see, benchmark points `A' to `D' correspond to `borderline' scenarios where the total $\bar{p}$ flux (i.e. the sum of the expected flux from DM and astrophysical background) is not significantly exceeding the data. Point `E', on the other hand, displays `how badly' the data is violated inside the excluded region.

The first apparent feature from Fig.~\ref{examples} is that one can actually exclude a small excess in anti-protons produced by relatively light Dark Matter particles because the \PAMELA\ data set have very small error bars at energies below 100 GeV, hence the strength of the constraints. 
It is then instructive to compare case `A' and `B': these two scenarios refer to the same DM mass and constraint procedure; they also predict a very similar flux, as can be seen in Fig.~\ref{examples}, but have a different annihilation cross section. The latter is  much larger for `A' than for `B'. This is because the propagation scheme was assumed to be `MIN' for the former and `MED' for the latter. With the `MIN' propagation set, the yield of anti-protons is about one order of magnitude smaller than with `MED' (since the galactic diffusion zone is much smaller in the former case) and therefore the constraint on the annihilation cross section is about one order of magnitude looser than for the `MED' case. On the other hand the constraint obtained for `MAX' (which is not shown here) is stronger than for `MED'. 

The comparison between points `B' and `C' shows the impact of the constraint procedures. Although both `B' and `C' have the same DM mass and propagation scheme, we find that the value of the annihilation cross section that is allowed for `C'  is larger than for `B'. The reason is that `C' corresponds to the scenario in which the limit is obtained by using the `marginalised background' procedure (i.e. where the background is allowed to retract within the uncertainties) so there is more room for DM while `B' corresponds to a `fixed background' scenario so the associated constraints are stronger.

Finally, the comparison between  `C' and `D' enables one to understand  why the `margi\-nalised background' constraints are rather independent of the DM mass, despite the fact that the error bars in the \PAMELA\ data become larger at larger energies. For a large DM mass (case `D') the $\bar{p}$ flux is shifted towards larger energies and rather negligible at $\sim$ 10 GeV with respect to the astrophysical background; there is thus little room to reduce the the bakcground (which alone has to fit the data at low energy) and consequently there is little room left for a DM contribution at large energies. As a result, the bound remains stringent.


\subsection{Generic bounds on $\sigma_{{\rm DM} \ {\rm DM}  \rightarrow W^+ W^-}$ from gamma-rays}
\label{gammaconstraints}

In DM scenarios, the $W^\pm$ production is associated with gamma-ray emission through (i) the decay and hadronisation of the decay products of the $W^{\pm}$ bosons, (ii) the radiation of a photon from the internal and/or final states associated with ${\rm DM} \ {\rm DM}  \rightarrow W^+ W^-$  (iii) DM annihilations into $\gamma\gamma$ and $\gamma Z$ (which can be seen as a higher order process based on ${\rm DM} \ {\rm DM}   \rightarrow W^+ W^-$). The first case leads to a {\it continuum} spectrum of $\gamma$-rays (the energy spectra can be e.g. found in~\cite{PPPC4DMID}, for any value of the DM mass); the second leads to {\it sharp features} in the $\gamma$-ray continuum spectrum and the third to {\it $\gamma$-ray lines}. The resulting fluxes from these process  have to be compared with the gamma-ray flux measurements from  the Milky Way or from other nearby galaxies. Therefore we now review the current $\gamma$-ray constraints derived in the literature (mainly from \Fermi-LAT analyses), paying particular attention to that derived from the $W^+ W^-$ channel since this is the main focus of our analysis.

\medskip 

\subsubsection{Continuum} The \Fermi-LAT collaboration has recently published two different analyses of the continuum diffuse gamma-ray emission from the Milky Way halo~\cite{FermiDiffuseMW,Ackermann:2012qk}. Since no clear DM signal has been found, these have been used  to set upper limits on the DM pair annihilation cross-section into various channels: e.g. $b\bar{b},gg,W^+W^-,e^+e^-,\mu^+\mu^-,\tau^+\tau^-$. For relatively light DM ($\sim$ 20 GeV) and e.g. the $b \bar{b}$ channel the limits reach the canonical value of the cross section (namely $\langle \sigma v \rangle =3\times 10^{-26} {\rm cm}^3/{\rm s}$), provided that the most aggressive procedure is used. For DM masses ${\cal O} (100)\ {\rm GeV}$ and for the $W^+ W^-$ channel the limit reads $\langle \sigma v \rangle \lesssim 2 \times 10^{-24} {\rm cm}^3/{\rm s}$. However the most stringent limits on the DM annihilation cross section have actually been obtained from another \Fermi-LAT analysis based on the diffuse $\gamma$-ray emission from dSph galaxies; these Dark Matter dominated objects indeed represent a good target for Dark Matter searches.  

\medskip 

In the present analysis we will use the results from~\cite{FermiDwarfs,GeringerSameth:2011iw} (see also~\cite{Cholis:2012am}). Although they use slightly different sets of targets \footnote{Ref.~\cite{GeringerSameth:2011iw} uses 7 dSphs $-$Bootes I, Draco, Fornax, Sculptor, Sextans, Ursa Minor, and Segue 1 while Ref.~\cite{FermiDwarfs} uses 10 dSphs $-$the same as above plus Carina, Coma Berenices and Ursa Major II$-$.}, slightly different datasets \footnote{Ref.~\cite{FermiDwarfs} uses 24 months between August 2009 and August 2010 while Ref~\cite{GeringerSameth:2011iw}) uses 3 years between August 2009 and August 2011.} and a different analysis procedure (\cite{GeringerSameth:2011iw} introduces a frequentist Neyman construction), they both derive consistent limits for the $b\bar{b}, \, W^+W^-,\mu^+\mu^-,\tau^+\tau^-$ channels. If we apply -- for definiteness -- the constraints from~\cite{FermiDwarfs} and assume a DM mass of $100\ {\rm GeV}$, the limit for the $W^+W^-$ channel reads as $\langle \sigma v \rangle < 8.5 \times 10^{-26} {\rm cm}^3/{\rm s}$.  The analysis procedure in~\cite{FermiDwarfs} allows one to incorporate the uncertainties associated with the DM energy density profile of individual dSph galaxies, which was shown to lead to an error band of about an order of magnitude on the constraint in~\cite{GeringerSameth:2011iw}. Here we do not attempt to address these issues; we simply draw the attention of the reader that these constraints have to be taken with care  until a better determination of the DM energy density profile in dSph galaxies is available. Consequently, we adopt the rather conservative constraints in this paper.

\medskip 

In Fig.~\ref{antip_crosssection} we compare the dSph galaxies limits with the \PAMELA\ anti-proton bounds that were derived in Sec.~\ref{pbarconstraints}. We see that, depending on the propagation scheme that has been chosen for the anti-protons, the dSph galaxies $\gamma$-ray bounds is somewhat more stringent or looser than the constraints from the anti-proton data. For example, for the `MED' case and `marginalized background', the $\bar p$ limits becomes more constraining than the $\gamma$-ray  bounds when $m_{\rm DM} \gtrsim 290$ GeV. However they are stronger than the $\gamma$-ray limits whatever the value of $m_{\rm DM}$ (assuming $m_{\rm DM}> 100$ GeV) for a `fixed' background. Since nevertheless the $\bar{p}$ and $\gamma$-ray limits are basically of the same order of magnitude,  we will include both constraints in our study.

\medskip

\subsubsection{Internal Bremsstrahlung and Final State Radiation}  Gamma rays produced directly as radiation from an internal line or a final state are in general suppressed by the fine structure constant, $\alpha$. However, for a $t$-channel diagram, the associated cross section can be enhanced when the intermediate particle is almost mass degenerated with the DM. Typically the enhancement factor is about $m_{\rm DM}^2/(M_I^2-m_{\rm DM}^2)$ where $M_I$ the mass of the intermediate particle (i.e. a chargino for neutralino pair annihilation into a $W^{\pm}$ pair). These process are model dependent and cannot be constrained generically but they will be included in our $\gamma$-ray estimates when we investigate the neutralino pair annihilations into $W^+ W^-$ in the pMSSM.

\medskip

\subsubsection{Line(s)}  Annihilations directly into 
$\gamma\gamma$ or $\gamma Z$ occur at one-loop level 
(since DM particles do not couple directly to photons) and are therefore generically suppressed. However they lead to a distinctive signature, namely a mono-energetic gamma-ray line at an energy $E=m_{\rm DM}$ or 
$E = m_{\rm DM} \, (1- m_Z^2/(4 m_{\rm DM}^2))$ which can be looked for.

With possible evidence for two gamma-ray lines at 129 and 111 GeV 
 (which have been speculated as originating from DM particles with a mass of about 130 GeV annihilating into $\gamma \gamma $ and $\gamma Z$), indirect detection of DM particles seem promising. Yet the existence of these lines remain to be confirmed by the \Fermi-LAT collaboration and their  origin to be shown as being exotic. Since the purpose of this study is to set constraints on the DM properties (and owing to these uncertainties on the existence and origin of these lines) we will disregard the results of \cite{line130GeV} and only consider the {\it constraints} which were reported by the \Fermi-LAT collaboration on line searches in the Milky Way~\cite{Ackermann:2012qk}, where the upper limits on $\sigma v_{{\rm DM} {\rm DM}  \rightarrow \gamma \gamma}$ and $\sigma v_{{\rm DM} {\rm DM}  \rightarrow Z \gamma}$  range from $0.03$ to $4.6 \times 10^{-27}{\rm cm}^3{\rm s}^{-1}$  and  $1$ to $10 \times 10^{-27}{\rm cm}^3{\rm s}^{-1}$ respectively, for DM masses up to 200 GeV. 
Constraints on $\sigma v_{{\rm DM} {\rm DM} \rightarrow \gamma \gamma}$ were also obtained from dSph galaxies~\cite{GeringerSameth:2012sr} but they are not as stringent as those obtained from the Milky Way.

Since the status of these searches is not definite, we made the choice to not include these constraints to perform the scans over the pMSSM parameter space. However we do check that the scenarios which survive the $\bar{p}$ and $\gamma$-ray constraints are not killed by these line searches.


\begin{figure}[h]
\begin{center}
\includegraphics[scale=0.6]{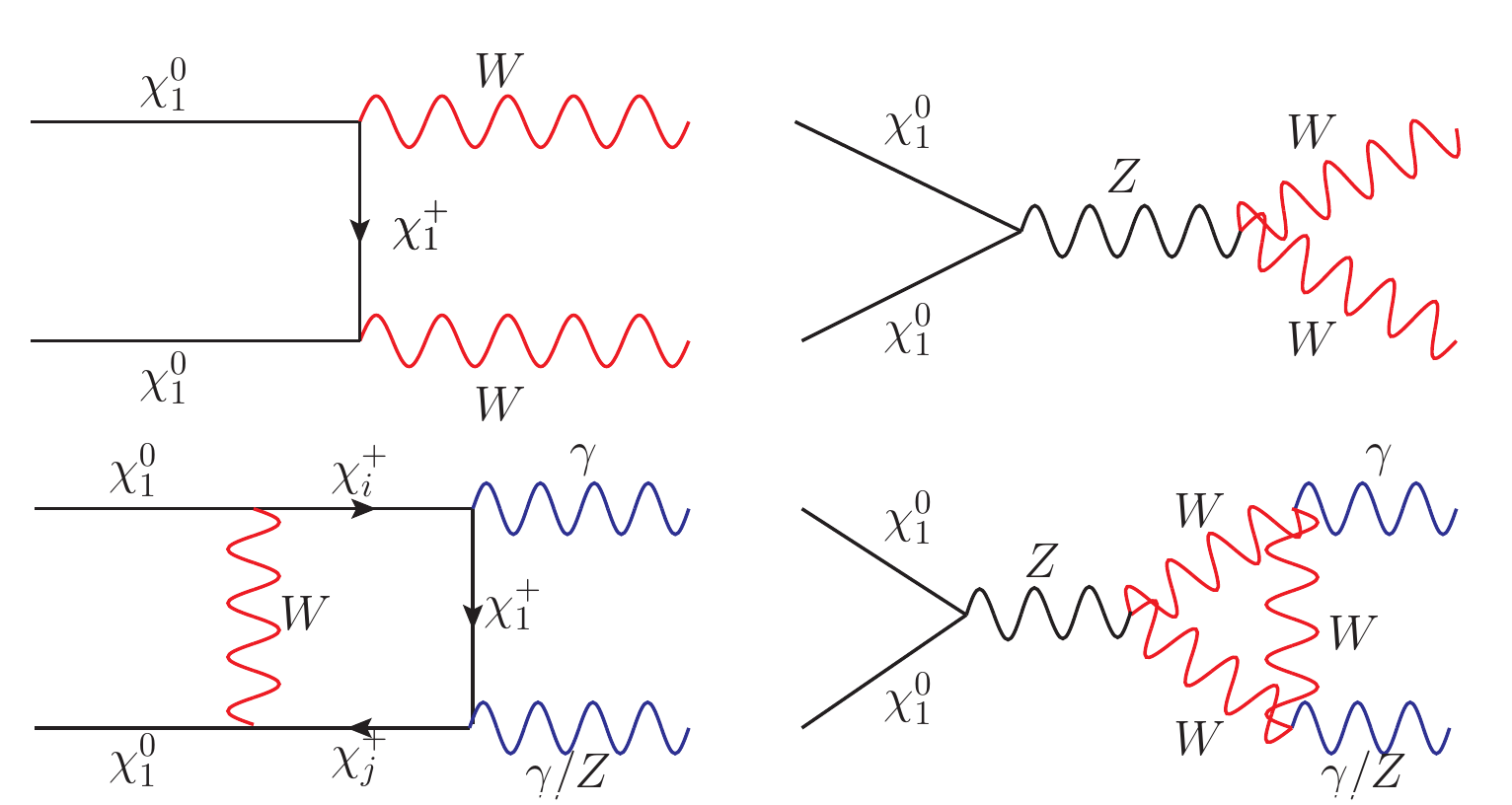}
\end{center}
\caption{\em \small {\bfseries Dominant neutralino pair annihilation diagrams} into $W^+ \, W^-$, $\gamma \, \gamma$ and $\gamma \, Z$ for this analysis. }
\label{pMSSM_diagrams}
\end{figure}

\section{Chargino-neutralino mass degeneracy}
\label{SuSy}

Now that we have obtained the maximal value of the Dark Matter pair annihilation cross section into $W^+ W^-$ that is observationally allowed as a function of the Dark Matter mass, we can focus on a specific Supersymmetric model and investigate the impact of this generic limit on the neutralino Dark Matter parameter space.


\subsection{ Neutralino pair annihilations into $W^+ W^-$}

In a scenario where all the sfermions are very heavy, the dominant neutralino annihilation channels are expected to be mostly into gauge bosons, more specifically into $W^+W^-$ pairs. All loop-induced $W^{\pm}$ production diagrams which involve sfermions are expected to be suppressed. Hence the process which are expected to lead to a significant $W^+ W^-$ production in the pMSSM only involve charginos and Z boson. The corresponding diagrams are displayed in Fig.~\ref{pMSSM_diagrams}. Since they correspond to $s-$ and $t-$channel diagrams, we typically expect resonant or enhanced annihilations when $m_{\chi^0_1} \sim  m_{Z}/2$ or $m_{\chi^0_1} \sim m_{\chi^\pm}$ (i.e. when the neutralino and chargino are mass degenerated). These ultimately enhance the neutralino pair annihilations into $\gamma \gamma$ \cite{Bergstrom:1997fh,Boudjema:2005hb} and $\gamma Z$ \cite{Boudjema:2005hb,Ullio:1997ke} through in particular the two `loop' diagrams displayed in Fig.~\ref{pMSSM_diagrams}.

The questions that we want to address in the next subsections are: 
i) which part of the SUSY parameter space is excluded by the $\bar{p}$ limits and do these limits exclude more allowed configurations than the $\gamma$-rays bounds? ii) which values of the neutralino-chargino mass degeneracy are actually constrained by astrophysical data?


\subsection{Exploring the supersymmetric parameter space}

To answer this, we will explore the pMSSM parameter space using the same Markov Chain Monte Carlo method as in \cite{Boehm:2012rh} coupled to the \texttt{micrOMEGAs} code \cite{Belanger} and the \texttt{SOFTSUSY} spectrum calculator \cite{Allanach}.

Our free parameters and their corresponding range are summarised in Table~(\ref{tab:range}). These  include  the soft mass terms associated with the squarks of the third generation (i.e. $M_{\tilde{Q}_3}$ and $M_{\tilde{u}_3}$) and the trilinear coupling $A_{t}$. To obtain sfermion masses at the TeV scale, we set all the soft masses to 2 TeV. In addition, we set the trilinear couplings to 0 and the CP-odd Higgs mass to 1 TeV.  In this framework, the bino mass $M_{1}$ does not exceed 500 GeV; our choice for the other parameters indeed ensures that the neutralinos and charginos are light and the mass splitting between the neutralinos and charginos remains relatively small. 

On top of these free parameters, we had to include some nuisance parameters over which we will marginalise~\cite{Dumont:2012ee}. These are related in particular to the quark content of the nucleons (since they have a non-negligible impact on the computation of the Dark Matter-nucleon scattering cross section) and the top mass (since it has an impact on the Higgs sector). All of them are  allowed to vary in the range [$N_{exp}$ -3$\sigma$, $N_{exp}$ +3$\sigma$], with $N_{exp}$ ($\sigma$) the corresponding experimental value (error), as shown in Table~\ref{tab:range}. 

We also require that the lightest Higgs mass only varies within the range allowed by the ATLAS and CMS experiments \cite{atlas:2012,cms:2012}, namely $m_h = 125.9 \pm 2.0$~GeV. However, by precaution, we checked that the scenarios which seemed allowed were compatible with the latest version of the HiggsBounds code~\cite{HiggsBounds-3.8.0} (even though the most recent LHC results on the Higgs~\cite{atlas:2012,cms:2012} are not included in this version).  Note that we did not add any requirement about the Higgs signal strength to perform the scans. Would ATLAS and CMS confirm  an `anomalous' Higgs signal strength into $\gamma \, \gamma$ (i.e. larger than SM expectations) with a high confidence level, the pMSSM would be difficult to reconcile with the data. However the principles of our analysis would remain valid and could still be used to constrain small mass degeneracies between the Dark Matter and another (e.g. $t-$channel exchange)  intermediate particle.

\begin{table}
\begin{center}
\begin{tabular}{|c|c|}\hline
Free parameter & Range \\ \hline \hline
$M_{1}$ & [10, 500] GeV\\ 
$M_{2}$ & [100, 1000] GeV\\ 
$\mu$ & [-2000, 2000] GeV\\ 
$\tan \beta$ & [2, 75]\\ 
$M_{\tilde{Q}_3}, M_{\tilde{u}_3}$ & [100, 3000] GeV\\
$A_{t}$ & [-8000, 8000] GeV\\ \hline
\end{tabular} 
\begin{tabular}{|c|c|c|}\hline
Nuisance parameter & Experimental value & Likelihood\\ \hline \hline
$m_u$/$m_d$ & 0.553 $\pm$ 0.043 \cite{Leutwyler:1996qg} & $\mathcal{L}_1(m_u/m_d, 0.51, 0.596, 0.043)$\\ 
$m_s$/$m_d$ & 18.9 $\pm$ 0.8 \cite{Leutwyler:1996qg} & $\mathcal{L}_1(m_s/m_d, 18.1, 19,7, 0.8)$\\ 
$\sigma_{\pi N}$ & 44 $\pm$ 5 MeV \cite{Thomas:2012tg} & $\mathcal{L}_1(\sigma_{\pi N}, 39, 49, 5)$\\ 
$\sigma_{s}$ & 21 $\pm$ 7 MeV \cite{Thomas:2012tg} & $\mathcal{L}_1(\sigma_{s}, 14, 28, 7)$\\ 
$m_t$ & 173.2 $\pm$ 0.9 GeV \cite{TEV} & $\mathcal{L}_1(m_t, 172.3, 174.1, 0.9)$\\ \hline
\end{tabular}
\caption{\em \small Range chosen for the {\bfseries pMSSM free parameters} and {\bfseries nuisance parameters}. \label{tab:range}}
\end{center}
\end{table}

The neutralino relic density is allowed to vary between $\Omega_{\chi^{0}_{1}} h^2 \in [1\%\textrm{ WMAP7}, \textrm{ WMAP7}]$ with $\Omega_{\rm WMAP7} h^2 = 0.1123 \pm 0.0035$,  using WMAP 7-year $+$ BAO $+$ $H_0$ and the \texttt{RECFAST 1.4.2} code~\cite{WMAP}. We do not consider smaller values of the relic density as these correspond to DM scenarios with very large values of the annihilation cross section and ultimately overproduce gamma-rays in the galaxy (i.e. are excluded) if their relic density is entirely regenerated, cf \cite{Williams:2012pz}.

For each scenario (corresponding to a point in the pMSSM parameter space), we then calculate the total likelihood function. The latter is a product of likelihood functions associated with each observable, nuisance parameters and free parameters which have been chosen according to the criteria described below.

\subsubsection{ $\mathcal{L}_1(x, x_{min}, x_{max}, \sigma)$}
To $m_h$, $\Omega_{\chi^{0}_{1}} h^2$ and all nuisance parameters, we associate a likelihood function  $\mathcal{L}_1$ which decays exponentially at the edges of a well-defined range $[x_{min}, x_{max}]$ with a variance $\sigma$:
\begin{eqnarray}
\mathcal{L}_1(x, x_{min}, x_{max}, \sigma)  = 
\begin{cases} 
e^{-\frac{\left( x - x_{min}\right)^2}{2\sigma ^2}} & \textrm{if} \: x < x_{min}, \\
 e^{-\frac{\left( x - x_{max}\right)^2}{2\sigma ^2}} &\textrm{if} \: x > x_{max}, \\
 1 & \textrm{for} \: x \in [x_{min}, x_{max}].
\end{cases}
\end{eqnarray}
Here $x$ is either $m_h$, the LSP relic density or the nuisance parameters. Note that we assume flat prior for all nuisance parameters. 
For the free parameters, we will consider a slight modification of the above function, namely 
\begin{eqnarray}
\mathcal{L}_1(x, x_{min}, x_{max}, \sigma) \, = 
\begin{cases} 
0 \ \textrm{for} \: x < x_{min} \ \textrm{or} \ x >  x_{max}, \\   
1  \ \textrm{for} \: x \in [x_{min}, x_{max}]
\end{cases}
\end{eqnarray}
so as to immediately reject all the scenarios in which one of the free parameters falls outside of the specified range.  In fact we also immediately reject points  where the neutralino is not the LSP, where the LEP limits on chargino, slepton and squark masses are not satisfied or the calculations of the SUSY spectrum fail. We did not implement LHC limits on sfermion masses because our requirement of a sfermion spectrum at the TeV scale should ensure that they are satisfied. However updates on direct searches for relatively light stop and sbottom would be useful to implement to further constrain the parameter space.

\subsubsection{$\mathcal{L}_2(x, x_{min}, x_{max}, \sigma)$}
 
We will use a Gaussian  Likelihood function, $\mathcal{L}_2$, for the ${\cal B}(b \rightarrow X_s^* \gamma)$ observable (one of the B-physics observables that we consider) with experimental mean value $\mu$ and theoretical $+$ experimental error $\sigma$ :
\begin{equation}
\mathcal{L}_2(x, \mu, \sigma) = e^{-\frac{\left(x-\mu\right)^2}{2\sigma ^2}}.
\end{equation}

These observables are important as they receive a potentially large contribution from chargino/stop loops when either one of these particles is light. This contribution can be compensated by the charged Higgs/top diagram but the latter is however suppressed when the charged Higgs mass is at the TeV scale.

\subsubsection{$\mathcal{L}_3(x, x_{min}, x_{max}, \sigma)$}
We also include a Likelihood function $\mathcal{L}_3$ for the 2012 XENON100 limits \cite{Aprile2012} to ensure that the scans do not select too large values of the Dark Matter-nucleon scattering cross section.  
In fact we also associate $\mathcal{L}_3(x, \mu, \sigma)$ to regions of the parameter space where ${\sigma v}_{{\chi^{0}_{1}\chi^{0}_{1}\rightarrow W^{+}W^{-}}} $ is greater than $10^{-27}$~${\rm cm}^3/{\rm s}$. The latter is defined as follows:
\begin{equation}
\mathcal{L}_3(x, \mu, \sigma) = \frac{1}{1+e^{-\frac{x-\mu}{\sigma}}}.
\end{equation}
 where the lower or upper experimental bound are associated with the  positive or negative variance $\sigma$ respectively. Note that some experimental measurements are very discrepant with the SM expectations (namely the anomalous magnetic moment of the muon $\Delta a_\mu$ and the branching ratio ${\cal B}(B^+ \rightarrow \tau^+ \bar \nu_\tau)$). These observables receive additional contributions from particles in the pMSSM but they are too small to explain the observations. Therefore we associate a Likelihood function to them which corresponds to $\mathcal{L}_3(x, \mu, \sigma)$ so that the Likelihood is equal to unity if the predictions are much below the measured value. 
The set of constraints that we use is summarised in Table~\ref{tab:constraints}. 

\renewcommand{\arraystretch}{1.3}
\begin{table*}[!tb]
\footnotesize{
\begin{tabular*}{1.031\textwidth}{|c|c|c|c|}
\hline 
Constraint & Value/Range & Tolerance & Likelihood \\ 
\hline \hline
       $m_h$ (GeV) \cite{atlas:2012,cms:2012} & [123.9, 127.9] & 0.1 & $\mathcal{L}_1(m_h, 123.9, 127.9, 0.1)$ \\ \hline 
       $\Omega_{\chi^0_1} h^2$ \cite{WMAP} & [0.001123, 0.1123] & 0.0035 & $\mathcal{L}_1(\Omega_{\chi^0_1} h^2, 0.001123, 0.1123, 0.0035)$ \\ \hline
       ${\cal B}(b \rightarrow X_s^* \gamma)$ $\times$ $10^{4}$ & 3.55 & exp : 0.24, 0.09 & $\mathcal{L}_2(10^{4} {\cal B}(b \rightarrow X_s^* \gamma), 3.55,$ \\ 
        \cite{Asner:2010qj,Misiak:2006zs} & & th : 0.23 &  $\sqrt{0.24^2 + 0.09^2 + 0.23^2})$\\ \hline
       $\sigma^{SI}_{\chi^0_1 {\rm Xe}}$ (pb) & ($m_{\rm DM}$, $\sigma_N$) plane & $\sigma_N(m_{\rm DM})$/100 & $\mathcal{L}_3(\sigma^{SI}_{\chi^0_1 {\rm Xe}}, \sigma_N(m_{\rm DM}), -\sigma_N(m_{\rm DM})/100)$ \\
       &  from \cite{Aprile2012} & & \\ \hline
       ${\sigma v}_{^{\chi^{0}_{1}\chi^{0}_{1}\rightarrow W^{+}W^{-}}}$ & 1 & 0.01 & $\mathcal{L}_3({\sigma v}_{^{\chi^{0}_{1}\chi^{0}_{1}\rightarrow W^{+}W^{-}}}, 1, 0.01)$ \\
       ($10^{-27}$~${\rm cm}^3/{\rm s}$) & & & \\ \hline
       $\Delta a_\mu$ $\times$ $10^{10}$ \cite{Davier:2010nc} & 28.70 & 0.287 & $\mathcal{L}_3(10^{10} \Delta a_\mu, 28.70, -0.287)$ \\ \hline 
       ${\cal B}(B_s \rightarrow \mu^+ \mu^-)$ $\times$ $10^{9}$~\cite{Aaij:2012ac} & 4.5 & 0.045 & $\mathcal{L}_3(10^{9}\  {\cal B}(B_s \rightarrow \mu^+ \mu^-), 4.5, -0.045)$ \\ \hline        
       $\Delta \rho$ & 0.002 & 0.0001 & $\mathcal{L}_3(\Delta \rho, 0.002, -0.0001)$ \\ \hline 
       $R_{B^+ \rightarrow \tau^+ \bar \nu_\tau} (\frac{\rm pMSSM}{\rm SM})$ \cite{Charles:2011va} & 2.219 & 2.219$\times10^{-2}$ & $\mathcal{L}_3(R_{B^+ \rightarrow \tau^+ \bar \nu_\tau}, 2.219, -2.219\times10^{-2})$ \\ \hline 
       $Z \rightarrow \chi^0_1 \chi^0_1$ (MeV) & 1.7 & 0.3 & $\mathcal{L}_3(Z \rightarrow \chi^0_1 \chi^0_1, 1.7, -0.3)$ \\ \hline 
       $\sigma_{e ^+ e ^- \rightarrow \chi^0_1 \chi^0_{2,3}} \times $ & 1 & 0.01 & $\mathcal{L}_3(\sigma_{e ^+ e ^- \rightarrow \chi^0_1 \chi^0_{2,3}} \times$ \\ 
       $ {\cal B}(\chi^0_{2,3} \rightarrow Z \chi^0_1)$ (pb) \cite{Abbiendi:2003sc} &&& $  {\cal B}(\chi^0_{2,3} \rightarrow Z \chi^0_1),1, -0.01)$ \\ 
\hline 
\end{tabular*}
}
\caption{\label{tab:constraints} \em \small {\bfseries Constraints imposed in the MCMC}, from \cite{Nakamura:2010zzi} unless noted otherwise.}
\end{table*}



\section{Results}
\label{results}

\begin{figure}[t]
	\parbox[b]{.48\linewidth}{
		\includegraphics[width=\linewidth]{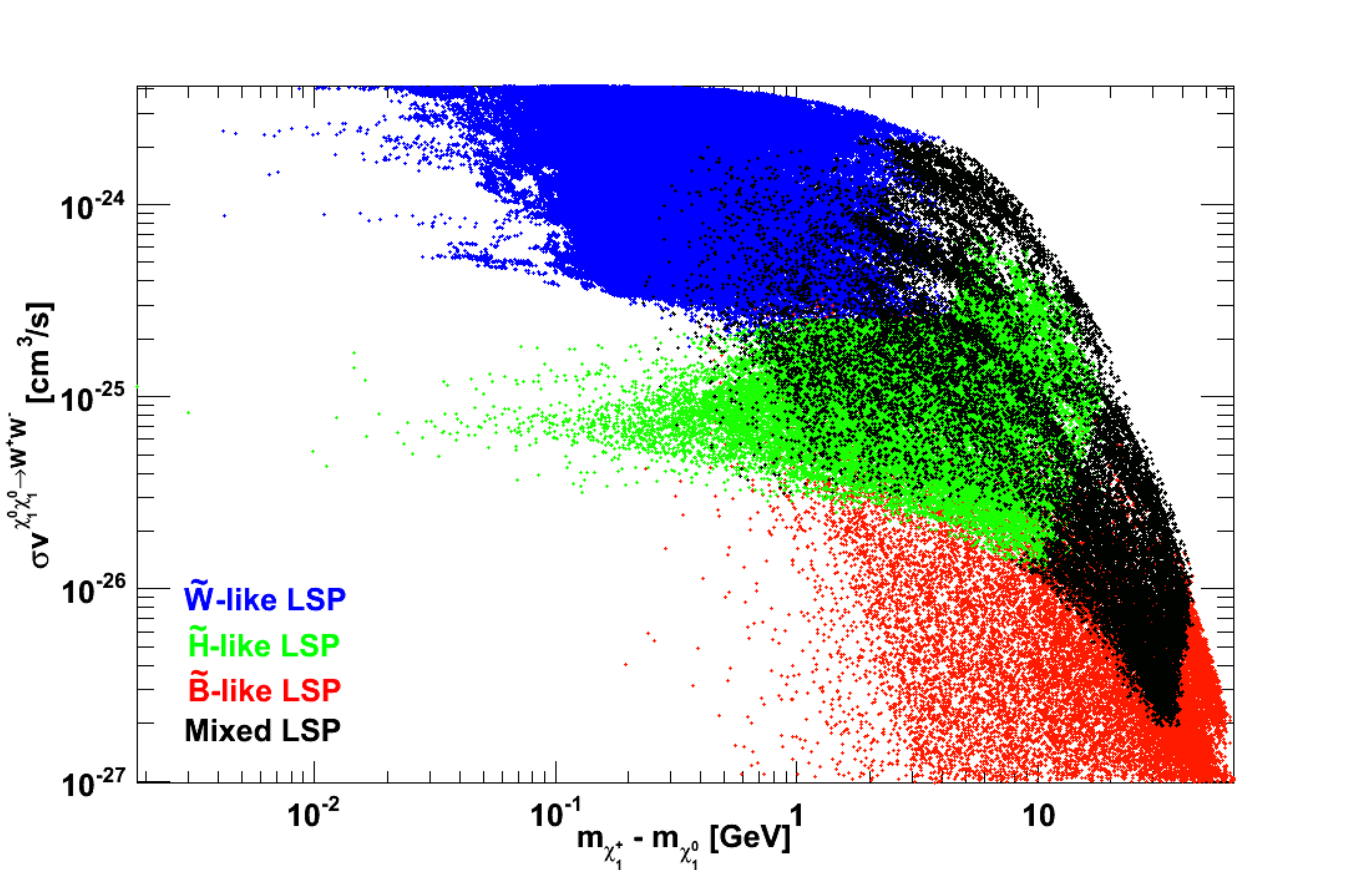}}
\parbox[b]{.48\linewidth}{
		\includegraphics[width=\linewidth]{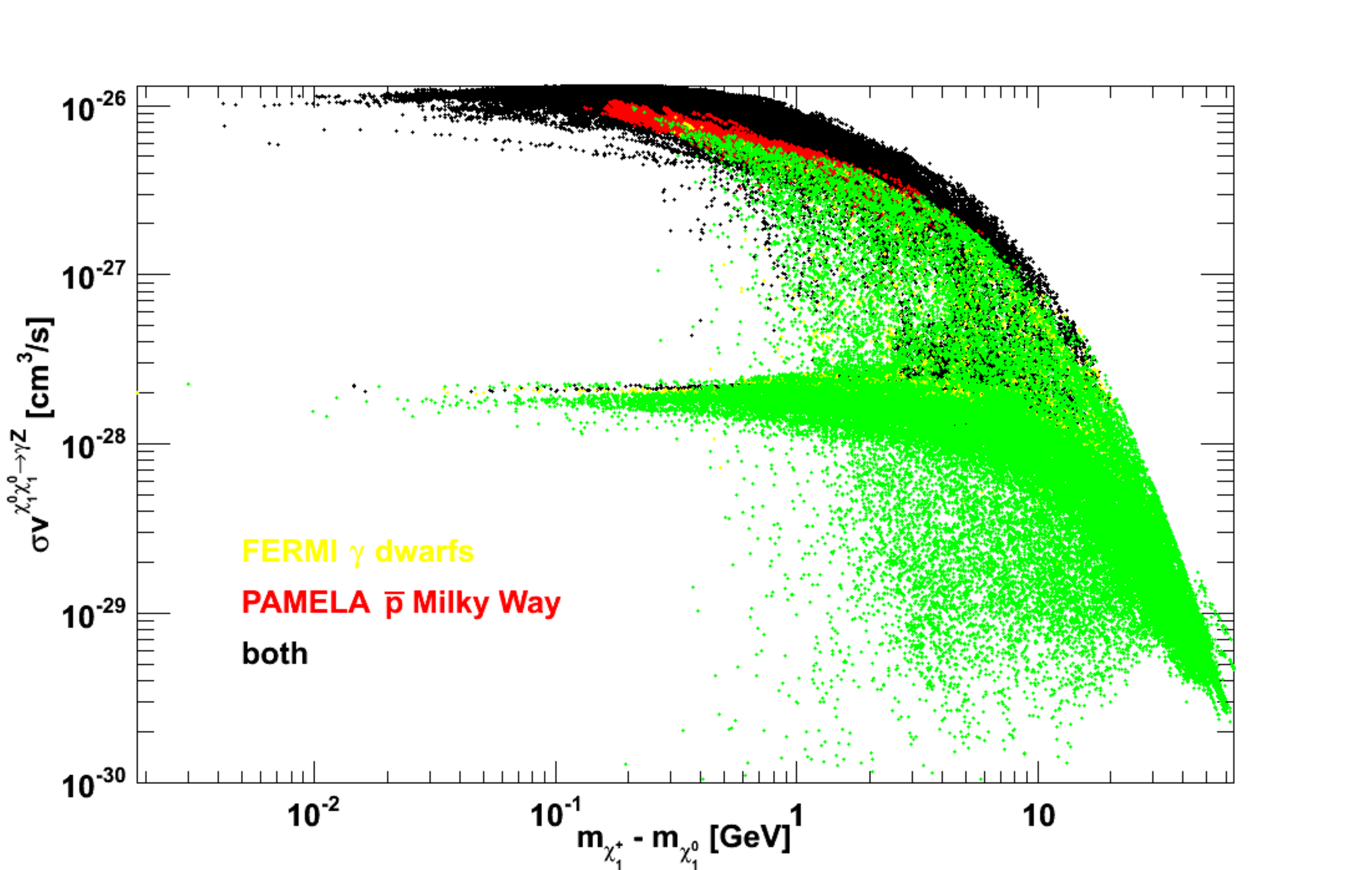}}
\parbox[b]{.48\linewidth}{
		\includegraphics[width=\linewidth]{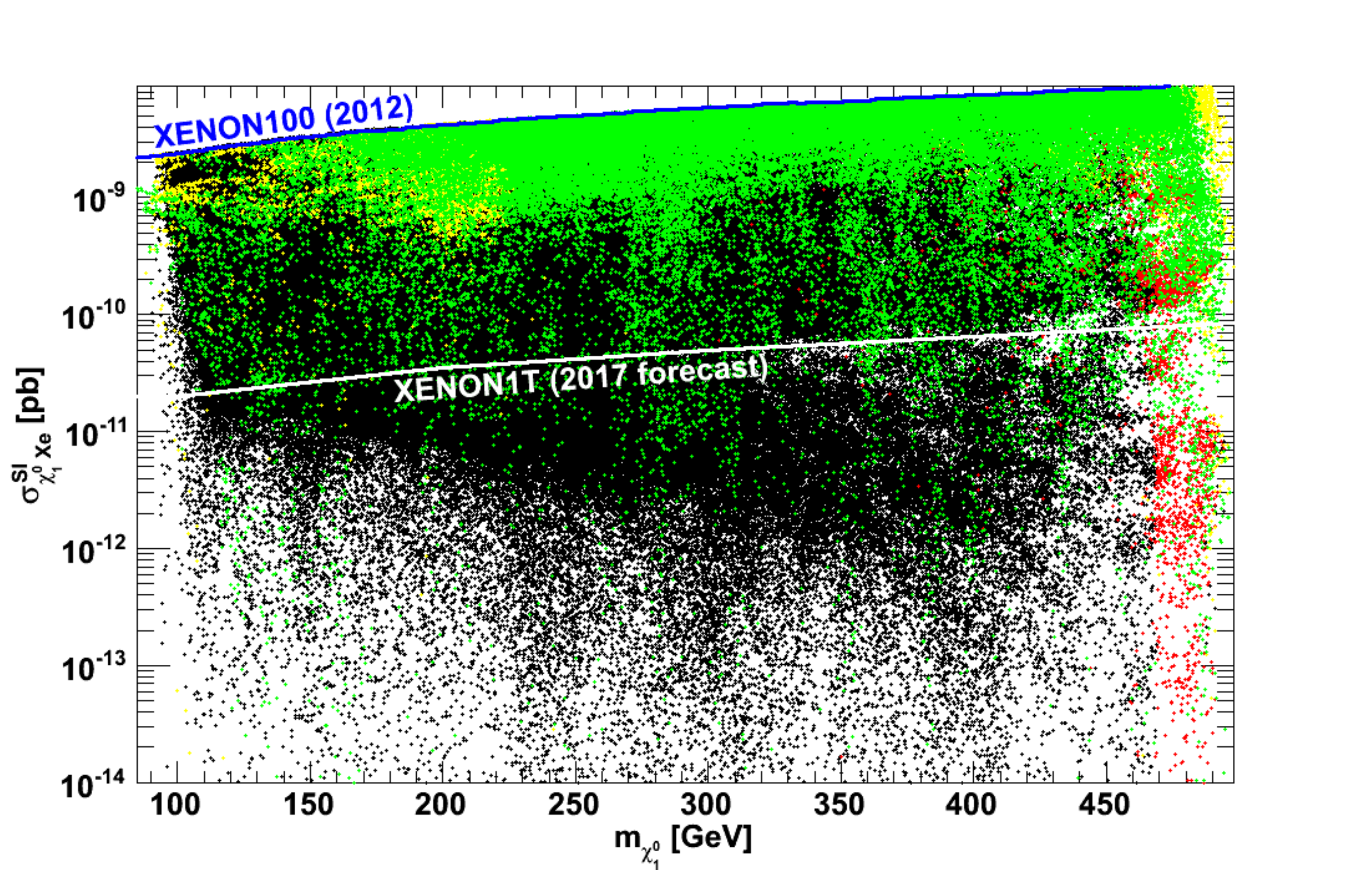}}
\parbox[b]{.48\linewidth}{	\quad \	
\includegraphics[width=\linewidth]{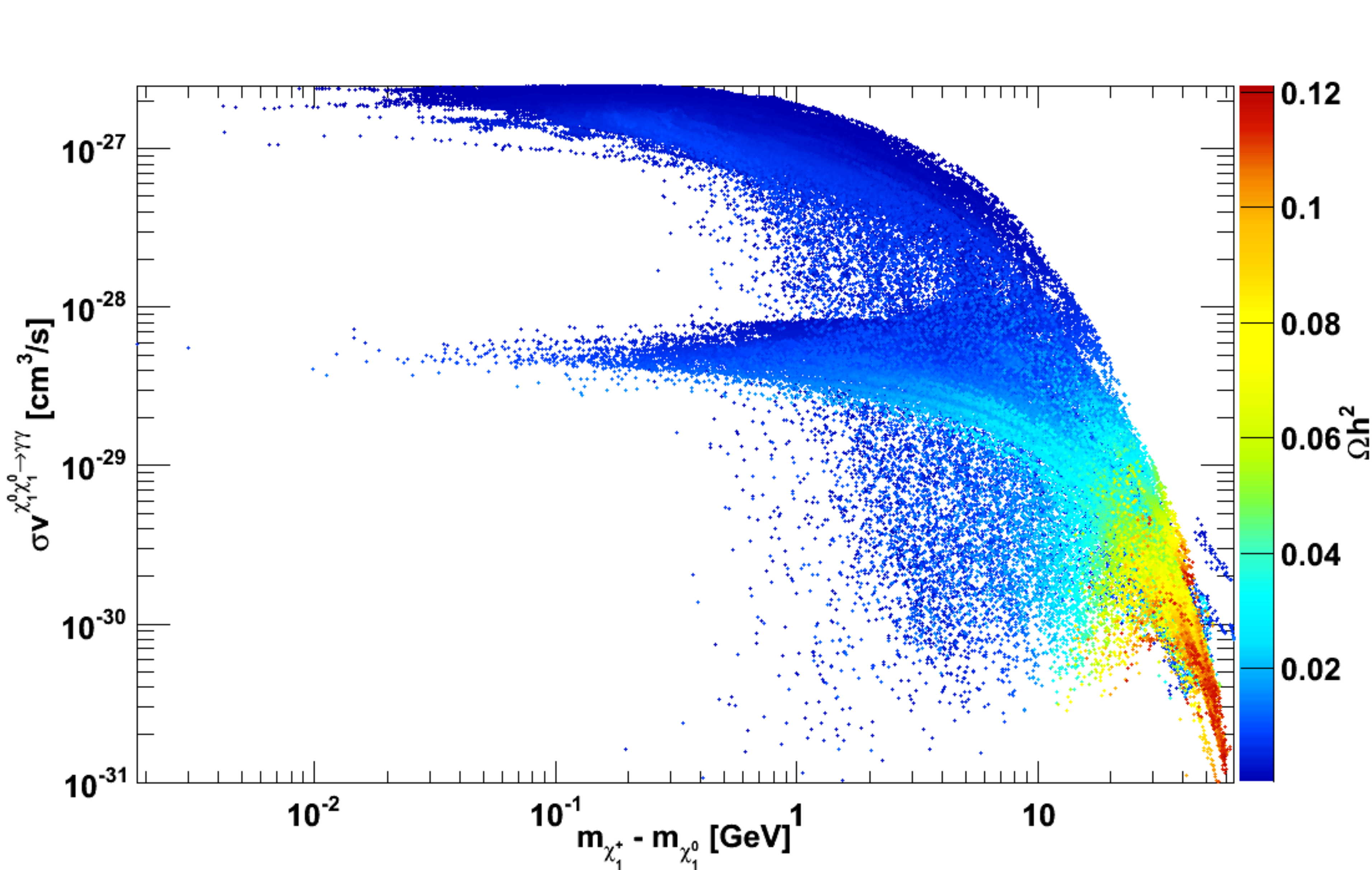}}
\caption{\em \small Plots of the neutralino pair annihilation cross section into $W^+ W^-$ (left upper panel) and $\gamma Z$ (right upper panel) as a function of the chargino-neutralino mass splitting and the spin-independent DM-nucleon cross section as a function of the Dark Matter mass (lower left panel) together with the XENON 2012 limit. The Freeze-Out relic density is displayed in the lower right panel for the annihilation cross section into $\gamma \gamma$ as a function of the mass splitting. }
\label{premierset}
\end{figure}

The results of our scans are shown in Fig.~\ref{premierset}. In the upper left panel is displayed the neutralino pair annihilation cross section into $W^+ W^-$ as a function of the mass degeneracy between the neutralino and the chargino and in terms of the neutralino composition.  In the upper and lower right panels we show the pair annihilation cross section into $\gamma Z$ and $\gamma \gamma$ respectively as a function of the neutralino-chargino mass degeneracy $\Delta m = m_{\chi_1^+} - m_{\chi_1^0}$  and in the lower left panel we give the forecasted  spin-independent elastic scattering cross section as a function of the neutralino mass for a Xenon-based experiment.

The left upper panel indicates the neutralino composition which maximises the $W^{\pm}$ production. As one can see scenarios where $\sigma v_{\chi^0_1 \chi^0_1 \rightarrow W^+ W^-}$ is the largest and the neutralino-chargino mass splitting is the smallest correspond to neutralinos with a very large wino fraction. Large values of both $\sigma v_{\chi^0_1 \chi^0_1 \rightarrow W^+ W^-}$ and the $\chi_1^0-\chi^+$ mass splitting correspond on the other hand to wino-dominated neutralinos but with a non negligible higgsino component. For these two types of wino-dominated configurations the neutralino and chargino mass degeneracy is large enough to make the $t-$channel (chargino) exchange diagram very large. As the wino fraction decreases, the mass splitting becomes larger and the $t-$channel chargino exchange diagram contribution decreases. However it remains large till the higgsino fraction which ensures large values of the $\chi_1^0-\chi^{+}-W^-$ coupling remains  significant (i.e. dominates over the bino fraction).

The upper right panel of Fig.~\ref{premierset} shows which values of the neutralino pair annihilation cross section into $Z \gamma$ are excluded by astrophysical data as a function of the neutralino-chargino mass splitting. A similar plot is shown for $\gamma \gamma$ but the colour code now illustrates the relation between the different values of this cross section and the neutralino `thermal' relic density. As one can see the shape of the scenario distribution for $\gamma \gamma$  and $Z \gamma$ is essentially the same in the $(\Delta m,\sigma v)$ plane. However  the $Z \gamma$ cross section is approximately 10 times larger than that for $\gamma \gamma$ for every scenario. Hence combining these two figures actually gives an information about the relic density of the scenarios which are excluded by astrophysical data.

In the $Z \gamma$ plot (upper right panel of Fig.\ref{premierset}), the points excluded by the \Fermi-LAT dSph continuum $\gamma$-ray data are displayed in yellow (we do not superimpose the constraints from line searches). Those correspond, by construction, to scenarios where there is a very large $W^{\pm}$ production (and thus a large contribution to the continuum $\gamma$-ray spectrum) but also to the few regions in which the LSP is heavy and where the $b\bar{b}$ final state (associated with the $s-$channel pseudo-scalar Higgs exchange and which cannot be discarded as it is significant) overproduces $\gamma$-rays. The regions which are excluded by the \PAMELA\ data are shown in red. The black points correspond to scenarios excluded by both the \PAMELA\ and  \Fermi-LAT data while those in green represent the points allowed by these two types of  constraints.

As one can see from the distribution of black points the largest  values of the annihilation cross sections into $Z \gamma$ (and therefore $W^+ W^-$) are excluded by both measurements. Since these scenarios correspond to a small (or relatively small) chargino-neutralino mass splitting and thus large values of the $t-$channel chargino exchange diagram, we can conclude that both \PAMELA\ and \Fermi-LAT data are relevant to constrain wino-dominated neutralinos. A small number of these configurations  is however constrained by only one of the \PAMELA\ or  \Fermi-LAT dataset but this does not affect the maximal value of the $\chi_1^0-\chi^+$ mass splitting that can be excluded by using  astrophysical considerations.

By inspecting where the neutralino pair annihilations into $Z \gamma$, $\gamma \gamma$ and $W^+ W^-$ are significant in these plots, one also finds that higgsino-dominated scenarios are constrained by both \PAMELA\ and \Fermi-LAT data because the box diagram (cf the lower left diagram in Fig.~\ref{pMSSM_diagrams}) still generates a large $W^{\pm}$ production. In fact, for such a LSP, the annihilation cross section into $ZZ$ also becomes non-negligible compared to that into $W^+W^-$. Since the expected $\gamma$-ray and $\bar{p}$ spectra from $W^{\pm}$ and $Z$ production are very similar, we accounted for them both when we made the comparison with the PAMELA\ and \Fermi-LAT data.

Finally the green points which pass all the constraints have a non-negligible bino component. This reduces the chargino exchange diagram contribution and thus enables to decrease the $W^{\pm}$ (and therefore anti-proton and $\gamma$-ray) production. For these bino-like configurations one expects the stop and sbottom exchange to be relevant, leading to quarks in the final state and possibly (in particular for $b \bar{b}$) an overproduction of gamma-rays. Note that such process would also compete with the neutralino pair annihilation into SM fermions near pseudo-scalar Higgs resonances for heavy neutralinos.

The lower left panel of Fig.\ref{premierset} indicates whether the spin-independent elastic scattering cross section is compatible with the latest results from the XENON100 experiment \cite{Aprile2012}. Again in green are the points which are astrophysically allowed, in black the points which are excluded by both \PAMELA\ and \Fermi-LAT data and in red or yellow the points which are either excluded by the \PAMELA\ or \Fermi-LAT experiments respectively. Clearly one can see that the combination of both the \PAMELA\ and \Fermi-LAT astrophysical constraints surpass the latest exclusion limit set by the XENON100 experiment. In fact in general the astrophysical constraints discussed in this paper even have a stronger exclusion power than the forecasted XENON1T limit, illustrating how important adding astrophysical knowledge is in this specific scenario.

Even though many configurations are excluded by the \PAMELA\ and \Fermi-LAT data, we do find scenarios which are neither excluded by the XENON100 2012 limit nor by the astrophysical constraints discussed in this paper.  
Hence the XENON100 experiment could still discover evidence for relatively light pMSSM neutralinos ($m_{\chi^0_1} < 500$ GeV) if these particles indeed exist. We note nevertheless that in \cite{Deb}, a constraint as strong as the XENON100 2012 limit was obtained by using the XENON100 2011 data and a Bayesian analysis where the full information available in the $(S_1,S_2)$ scintillation plane was exploited. It is therefore likely that the XENON100 experiment can improve its present exclusion limit with the 2012 data and rule out some of the configurations shown here in green.  

In these figures we have assumed that the relic density was regenerated  at 100 $\%$  for candidates with a total  annihilation cross section much larger than the 
`thermal' one (i.e. with a suppressed Freeze-Out relic density). This way we could ensure a fair comparison between theoretical expectations and the limits set by the \Fermi-LAT and XENON100 experiments.  Looking at the $\sigma v_{\chi^0_1 \chi^0_1  \rightarrow \gamma \gamma}$ plot, one sees that invoking  regeneration is needed for all scenarios with a chargino-neutralino mass splitting smaller than $\sim$ 20 GeV \footnote{For larger values of the mass splitting, no regeneration assumption is required but the annihilation cross sections into $\gamma \gamma$ and $\gamma Z$ are strongly suppressed. In particular $\sigma v_{\chi^0_1 \chi^0_1  \rightarrow \gamma \gamma}$ is much below $10^{-29} \ \rm{cm^3/s}$.}.  Assuming that all these candidates have the correct relic density, we could indeed exclude scenarios with a neutralino-chargino mass splitting up to 20 GeV and values of $\sigma v_{\chi^0_1 \chi^0_1 \rightarrow \gamma Z}$ down to $10^{-28} \ \rm{cm^3/s}$ (see Fig.\ref{premierset}), corresponding to $\sigma v_{\chi^0_1 \chi^0_1 \rightarrow W^+ W^-} > 10^{-25} \ \rm{cm^3/s}$ and $\Omega h^2 \ll 0.06$.
However, relaxing the regeneration assumption would completely relax the exclusion regions and therefore the bound on the mass splitting (apart perhaps from scenarios with extremely small mass splitting).

As a side comment regarding the so-called `130 GeV line':  we do  find scenarios where $\sigma v_{\chi^0_1 \chi^0_1 \rightarrow \gamma \gamma} \simeq  10^{-27} \ \rm{cm^3/s}$, which is the value of the cross section that is required to explain the feature in the spectrum. These configurations predict a neutralino-chargino mass splitting greater than $\sim 0.2$ GeV.  However none of the points corresponding to neutralinos with a mass of about 130 GeV are allowed by the \PAMELA\ data. Hence, our results suggest that one cannot explain the `130 GeV line' in our simplified version of the pMSSM, which is in agreement with~\cite{Cohen:2012me,Buchmuller:2012rc}. Indeed, due to the anti-proton limit, scenarios with $\sigma v_{\chi^0_1 \chi^0_1  \rightarrow \gamma \gamma} \simeq  10^{-27} \ \rm{cm^3/s}$ rather correspond to neutralinos with a mass of about 400-450 GeV. In fact, for the same reason, all the points with $\sigma v_{\chi^0_1 \chi^0_1  \rightarrow \gamma \gamma}>2 \times 10^{-28} \ \rm{cm^3/s}$ correspond to configurations where $m_{\chi^0_1} >$200 GeV. Finally note that in the pMSSM the existence of 130 GeV neutralinos should give rise to a second $\gamma$-ray line at $\sim$ 111  GeV (on top of that at 130 GeV),  corresponding to the neutralino pair annihilation into $\gamma Z$. Given our prediction for $\gamma Z$ and $\gamma \gamma$, the flux associated with this 111 GeV line should be about ten times larger than that corresponding to the 130 GeV line, which is in conflict with the observations.

\begin{figure}[t]
\includegraphics[width=\textwidth]{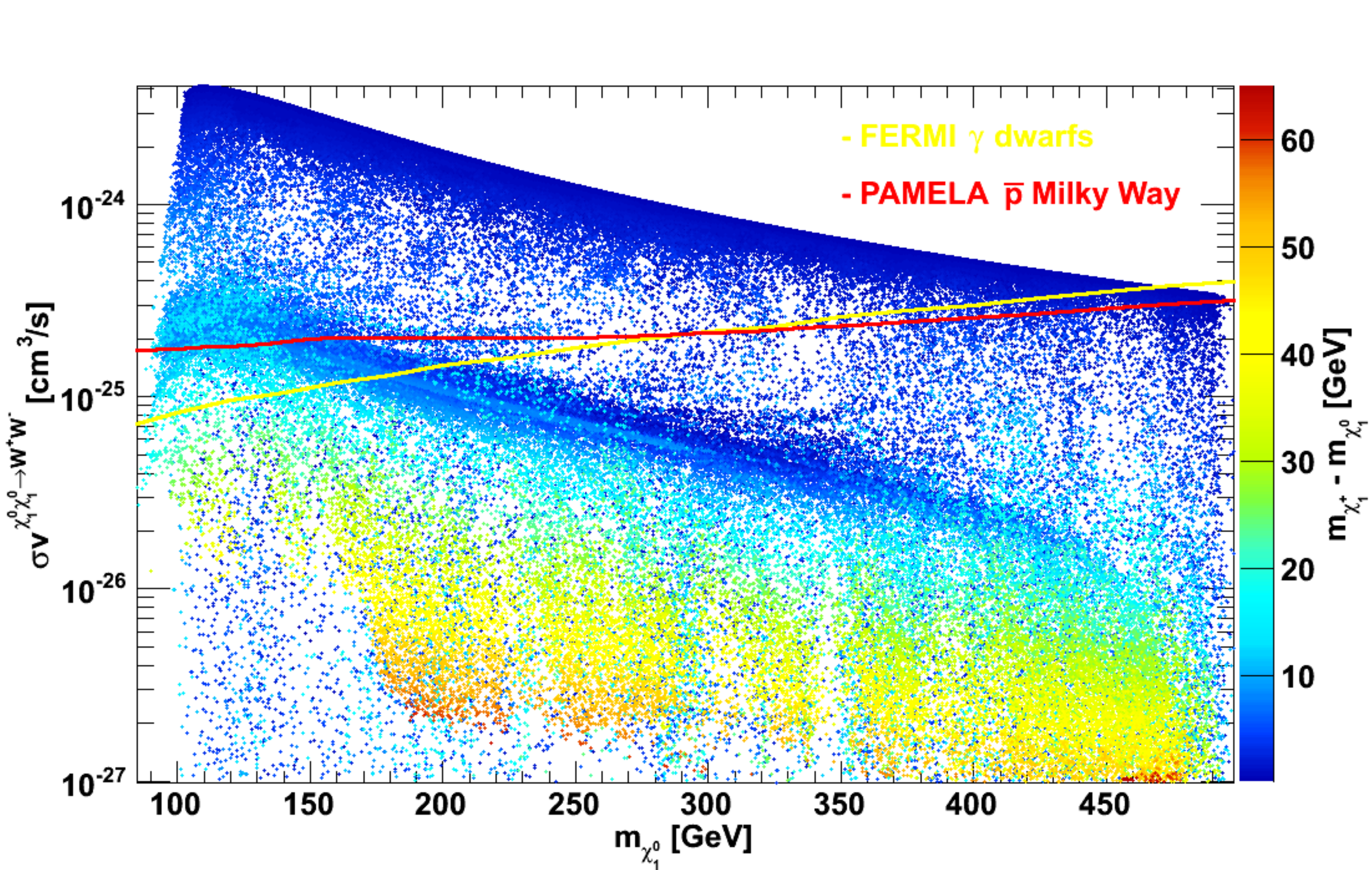}
\caption{\em \small  {\bfseries Annihilation cross section 
into $W^+ W^-$} as a function of the neutralino Dark Matter mass. The chargino NLSP-neutralino LSP mass splitting is shown as colour code.}
\label{WW}
\end{figure}

Finally,  in Fig.\ref{WW}, we show the annihilation cross section 
into $W^+ W^-$ as a function of the neutralino mass and superimpose the \PAMELA\ (for the `MED' set of propagation parameters and marginalised background, i.e. the conservative limits) and \Fermi-LAT limits (red and yellow lines respectively). The colour code indicates the different values of the neutralino-chargino mass splitting. As can be seen from this plot, the \PAMELA\ and \Fermi-LAT constraints are actually complementary. The \Fermi-LAT limit excludes more configurations below 300 GeV than the \PAMELA\ bound but it assumes that the observations are independent of the Dark Matter energy distribution in dSph galaxies, which can be debated~\cite{Hooper:2003sh,Charbonnier:2011ft}. In contrast, the anti-proton limit excludes a bit more configurations than the gamma-rays  above 300 GeV. This is reassuring since it is set by observations `within' the galaxy but the drawback is that it relies on a specific choice of propagation parameters and astrophysical knowledge of astrophysical sources.  In any case, the fact that both limits exclude similar configurations  enables us to validate the exclusion region that we found. 

Hence the main information that one can read of from this plot, combined with that displayed in Fig.~\ref{premierset}, is that:
\begin{itemize} 
\item one can rule out neutralino-chargino mass splitting up to $\sim$ 20 GeV if $m_{\chi_1^0} \lesssim 150$ GeV and the neutralino is a mixture of wino and higgsino 
\item one can exclude all scenarios in which the wino-chargino mass difference is smaller than 0.2 GeV for $m_{\chi^0_1} < 500$ GeV, thanks to both \PAMELA\ and \Fermi-LAT data. 
\end{itemize}

\section{Conclusions}
\label{conclusions}

\medskip
In this paper we explicitly derived the constraints on the ${\rm DM} {\rm DM} \rightarrow W^+ W^-$ annihilation cross section by using the \PAMELA\ anti-proton data and paying particular attention to the choice of propagation parameters and uncertainties on the astrophysical background. Our results are independent of the so-called \PAMELA\  positron excess and are obtained for two different (fixed vs marginalised) choices of the background spectrum; they are also consistent with the enhancement factor which was derived in \cite{salati} and the detailed analysis of \cite{Evoli:2011id}, for the cases where the propagation parameters overlap.

\medskip 
We then compared these bounds with the most stringent gamma-ray limits which have been derived using the \Fermi-LAT\ measurements of the gamma-ray continuum spectrum from dSph galaxies, for the same DM annihilation channel and DM mass range. We found that the anti-proton constraints appear to be very competitive with the gamma-ray bounds. More precisely, choosing the `MED' propagation scheme, the $\bar p$ constraints are slightly weaker than the $\gamma$-ray ones when $m_{\rm DM} \lesssim  300$ GeV and slightly stronger for $m_{\rm DM} \gtrsim 300$ GeV.  On the other hand, the anti-proton constraints are stronger if we assume the 'MAX' set of propagation parameters and less powerful if we assume the `MIN' set. We also recall that the gamma ray limits themselves may be subject to some uncertainties related to the modelling of the DM profile in dSph  galaxies. 

\medskip
Finally we applied as fiducial limits the $\bar p$ constraints relative to `MED' and the marginalized astrophysical background to the neutralino LSP in a simplified version of the pMSSM, where we set all the sfermion masses (apart from that of the third generation) to the TeV scale. We found that the fiducial \PAMELA\ anti-proton and \Fermi-LAT gamma-ray limits rule out small but non negligible neutralino-chargino mass splittings. In particular for $m_{\chi_1^0} \lesssim 150$ GeV, one can rule out mass splittings up to 20 GeV. Our results also suggest that pure wino or wino-like neutralinos are excluded if they are lighter than 450 GeV. Overall, this limit surpasses the bounds that can be set by using the XENON100 data and even in fact than the projected XENON1T limit. 

\medskip
Hence from this work, we conclude that present indirect detection data already enable one to exclude regions of the parameter space where the neutralino-chargino mass splitting is small but non negligible. Since these regions are difficult to probe directly at the LHC, these findings show that \Fermi-LAT and \PAMELA\ data constitute modern tools to explore the supersymmetric parameter space and even beat LHC (and also in fact Direct Detection) searches on their own territory, even though -- on the negative side -- 
they assume a regeneration of the relic density for neutralinos with a very large annihilation cross section.


\section*{Acknowledgment}
We thank the Galileo Galilei Institute for Theoretical Physics in Florence for the hospitality and the INFN for partial support during the completion of this work.  
The work of MC is supported in part by the French national research agency ANR under contract ANR 2010 BLANC 041301 and by the EU ITN network UNILHC.
The work of AP was supported by the Russian foundation for Basic Research, grant RFBR-10-02-01443-a and the LIA-TCAP of CNRS.
JDS is supported by the CMIRA 2011 EXPLO'RA DOC grant and would like to thank IPPP for its hospitality.

\end{document}